\documentstyle[twoside,fleqn,espcrc2,epsf]{article}

\title{Evidence for pronounced quark loop effects in QCD}

\author{Robert D.~Mawhinney\address{Department of Physics,
  Columbia University, New York, NY 10027, USA}\thanks{The 2 flavor
  calculation reported here was done at Columbia in collaboration with
  Shailesh Chandrasekharan, Norman H. Christ, Dong Chen, Weonjong Lee,
  and Decai Zhu; The 4 flavor calculations were done at Columbia in
  collaboration with Dong Chen, Yubing Luo and Norman H. Christ.  The
  quenched calculations were done at Columbia and the Ohio State
  University, in collaboration with Norman H. Christ, Dong Chen and
  Gregory W.~Kilcup (of OSU). This work was supported in part by the
  Department of Energy.  Presented at Lattice QCD on Parallel Computers,
  Tsukuba, Japan, March 1997.}
}

\begin{document}

\def\thepage{CU--TP--839}
\thispagestyle{myheadings}

\begin{abstract}
We have measured the hadron spectrum in lattice QCD, using staggered
fermions, for 0 (the quenched approximation), 2 and 4 light degenerate
dynamical quarks.  In addition to earlier results involving
extrapolations in valence quark masses for fixed dynamical mass, we
also report results where we extrapolate in the dynamical mass for 4
flavors.  We see a marked difference in the hadron spectrum for 2 and 4
flavors;  the hadron spectrum is nearly parity doubled for 4 flavors,
indicating smaller effects of chiral symmetry breaking.  This
pronounced effect in the hadron spectrum cannot be removed by a simple
change in scale as the number of light quark flavors is changed.
Further simulations at larger volume are needed to rule out finite
volume effects.

\end{abstract}

\maketitle

\section{INTRODUCTION}

During the past two years, the lattice QCD group at Columbia has
been exploring the effects of light (by today's standards) dynamical
quarks on the hadron spectrum in zero temperature simulations of
QCD.  Light quark effects are readily seen at finite temperature,
where studies have shown the order of the QCD phase transition is
sensitive to the number of light quarks, as expected from theoretical
arguments.  Zero temperature simulations to date, including
the ones reported here, have shown little difference between
0 and 2 flavors of dynamical quarks.  To see if this continues,
we have done extensive simulations using 4 flavors ($N_f = 4 $).

On our existing 256-node, 16 Gflop peak speed computer, we are
essentially limited to lattices of size $16^3 \times N_t$ for light
quark mass simulations ($ ma \sim 0.01$, where $m$ is the bare
staggered quark mass).  Using $m_\rho$ to set the scale gives us
lattice spacings of $a^{-1} \sim$ 2 GeV, for the lighter masses we
simulate.  (The exact value for the scale depends on whether one uses
$m_\rho$ extrapolated to zero quark mass or not.)  This also makes our
lattice about 1.5 fermi in the spatial directions.  Quenched
simulations show this introduces errors at the 5-10\% level
\cite{Gottlieb97} and similar systematic errors can be expected for the
full QCD simulations, although complete studies await the arrival of
more powerful computers.  Since our volume is limited, we cannot
extrapolate to the infinite volume limit and also not the continuum
limit.  Thus, our results are expected to be changed quantitatively as
these limits are taken, but there is no compelling reason to expect a
qualitative change.

\section{SIMULATIONS}

Table~\ref{tab:parameters} lists the parameters for our simulations.
It is important to note that the 2 flavor results employ an inexact
algorithm \cite{Gottlieb87}, with errors proportional to $( \Delta \tau
)^2$.  Since we have used the same step size, for a given dynamical
mass, in both the 2 and 4 flavor calculations and the 4 flavor
calculations have a very high acceptance, the finite step size errors
for the inexact algorithm are apparently very small.  We have used a
smaller conjugate gradient stopping condition for the 4 flavor runs,
since this enters into the accept/reject step, and we wanted to insure
that this step is accurate.

\begin{table*}[htb]
\caption{Run parameters for the simulations whose results are reported
here.  The 2 flavor, $ma = 0.015$, 0.02 and 0.025 results
are from [3]. The longer thermalization time for the 0.025 run
merely reflects measurements not being made from the beginning of
the run.}
\label{tab:parameters}
\begin{center}
\begin{tabular}{|l|c|c|c|c|} \hline
    	& $m_{\rm dyn}a = 0.01 $ & $m_{\rm dyn}a = 0.015 $
	& $m_{\rm dyn}a = 0.02 $ & $m_{\rm dyn}a = 0.025 $\\ \hline
\hline
\multicolumn{5}{|l|}{$N_f = 2$, $\beta = 5.7 $ } \\ \hline
  volume		& $16^3\times40$	& $16^3\times32$ 
			& $16^3\times32$	& $16^3\times32$\\\hline
  run length		& 4870	& 3010	& 1425 & 2830 \\ \hline
  thermalization	& 250	& 250	& 250	& 530 \\ \hline
  step size		& 0.0078125	& 0.0078125 & 0.01 & 0.01 \\\hline
  CG stopping condition	& $1.01\times10^{-5}$	& $1.13\times10^{-5}$
			& $1.13\times10^{-5}$	& $1.13\times10^{-5}$
			\\\hline
\hline
\multicolumn{5}{|l|}{$N_f = 4$, $\beta = 5.4 $ } \\ \hline
  volume		& $16^3\times32$	&
			& $16^3\times32$	& \\\hline
  run length		& 4450	& & 2725 & \\ \hline
  thermalization	& 250	& & 250	& \\ \hline
  step size		& 0.0078125	& & 0.01 & \\ \hline
  CG stopping condition	& $1.13\times10^{-6}$	& 
  			& $1.13\times10^{-6}$	& \\ \hline
  acceptance 		& 0.95 & & 0.99 & \\ \hline
\end{tabular}
\end{center}
\end{table*}

In these simulations we have measured hadron correlators for a variety
of different source types and sizes with many different valence quark
masses (the masses which appear in the explicit Green functions for the
hadron).  This allows us to fit for excited states and check the
stability of our effective mass plateaus \cite{dch}.  We also have
measured the chiral condensate for valence quark masses spanning 10
orders of magnitude which is useful as a probe of finite volume effects
and chiral symmetry breaking.

\section{PARTIALLY QUENCHED RESULTS}

Results from some of these earlier simulations can be found in
\cite{dch}, \cite{dch-rdm} and \cite{rhic96}.  We studied the
valence hadron spectrum for 0, 2 and 4 quark flavors with a
single fixed dynamical quark mass $m_{\rm dyn}a = 0.01$ for
$N_f = 2$ and 4.  For $N_f = 2 $ we used $\beta = 5.7 $ and for
$N_f = 4$, $\beta$ was 5.4.  The parameters were chosen so that the
rho mass, extrapolated to zero valence quark mass, was equal
(we achieved equality at the 2\% level) for 0, 2 and 4 flavors.
In choosing our parameters for the 4 flavor simulation, we were
helped by the earlier 4 flavor work of \cite{mtc}. 

\pagenumbering{arabic}
\addtocounter{page}{1}

\begin{figure}[hbt]
\epsfxsize=\hsize
\epsfbox{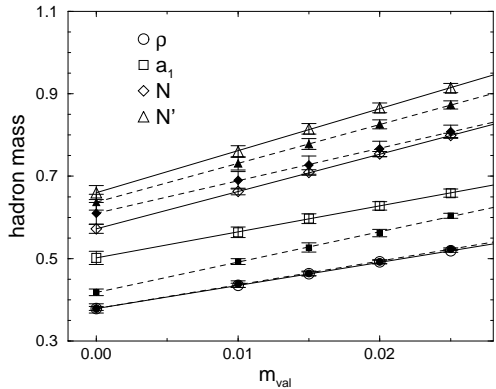}
\caption{Masses for the parity partners, $\rho$ and $a_1$, $N$ and
$N'$ for a partially quenched calculation.  Unfilled symbols are for
2 flavors, filled circles are for
4 flavors.  All simulations had  $m_{\rm dyn}a$ fixed at 0.01.}
\label{fig:parity_partners_val}
\end{figure}

Figure \ref{fig:parity_partners_val} shows valence hadron masses
for partially quenched 2 and 4 flavor simulations with
$m_{\rm dyn}a = 0.01$.  The parity partners $\rho$, $a_1$ and
$N$, $N'$ for 4 flavors show less splitting, particularly
in the $m_{\rm val}a \rightarrow 0 $ limit, than for 2 flavors.

The other major results of this work were (using $m_\rho$ to set the
scale):
\begin{enumerate}
\item
  $m_N$ and $m_\pi$ are very similar between 0 and 2 flavors.
  However, the mass of the parity partners, $m_{N'}$, $m_{a_1}$, differs
  by 2-3$\sigma$ between these two cases, with $m_{N'}$ and $m_{a_1}$
  closer to their parity partners for 2 flavors.
\item
  For 4 flavors, $m_N$/$m_\rho$ extrapolated to zero valence quark
  mass is 7\% larger for 4 flavors than for 2 flavors
  (about 2$\sigma$).  The dependence of this
  quantitative difference on lattice volume and spacing deserves
  further study.
\item
  Linear fits of $m_\pi^2$ to the valence quark mass do not go to
  zero as expected for a Goldstone particle.  This is consistent
  with the finite volume effects expected from a cutoff in the
  Dirac eigenvalue spectrum.  In particular, the mass of the
  Goldstone particle at zero quark masses is predicted to increase
  as the amount of chiral symmetry breaking decreases and this
  is seen in the data \cite{rhic96}.
\end{enumerate}

These results agree with the naive expectation that adding more powers
of the determinant (more quark flavors) suppresses the small eigenvalues
of the Dirac operator, which decreases the effects of chiral
symmetry breaking since the chiral condensate is given directly by the
density of eigenvalues at the origin.  The size of the effect, for the
quark masses we are currently able to simulate with, is certainly
larger than anticipated, given the close agreement between the 0 and 2
flavor simulations.

\section{FULL QCD RESULTS}

Having seen a marked decrease in the splitting between parity
partners in the partially quenched hadron spectrum, we have been
investigating the hadron spectrum in full QCD as a function of $N_f$.
To date, we have results for two dynamical mass values for
4 flavors and we can compare these to four dynamical mass values for
2 flavors.  The three heavier masses for 2 flavors are results
from \cite{hdm91} while the $m_{\rm dyn}a = 0.01$ is newer and includes
3 times the data of the 0.01 point in \cite{hdm91}.  Table
\ref{tab:masses} gives some of the hadron masses for our 2 and 4
flavor simulations.

\begin{table*}[htb]
\caption{ Hadron masses for 2 and 4 flavor QCD.}
\label{tab:masses}
\begin{center}
\begin{tabular}{|l|l|l|l|l|} \hline
    	& $m_{\rm dyn}a = 0.01 $ & $m_{\rm dyn}a = 0.015 $
	& $m_{\rm dyn}a = 0.02 $ & $m_{\rm dyn}a = 0.025 $\\ \hline
\hline
\multicolumn{5}{|l|}{$N_f = 2$, $\beta = 5.7 $ } \\ \hline
$\pi$	& 0.249(2) & 0.293(2) & 0.349(2) & 0.388(1) \\ \hline
$\rho$  & 0.435(5) & 0.455(8) & 0.501(7) & 0.551(4) \\ \hline
$a_1$   & 0.564(13)& 0.594(22)& 0.638(18)& 0.755(32) \\ \hline
$N$     & 0.663(8) & 0.685(10)& 0.781(10)& 0.839(6) \\ \hline
$N'$    & 0.760(14)& 0.833(38)& 0.905(46)& 1.022(41) \\ \hline
\hline
\multicolumn{5}{|l|}{$N_f = 4$, $\beta = 5.4 $ } \\ \hline
$\pi$   & 0.292(5) &          & 0.357(3) &           \\ \hline
$\rho$  & 0.438(8) &          & 0.499(6) &           \\ \hline
$a_1$   & 0.493(7) &          & 0.617(14)&           \\ \hline
$N$     & 0.690(21)&          & 0.773(10)&           \\ \hline
$N'$    & 0.731(16)&          & 0.872(13)&           \\ \hline
\end{tabular}
\end{center}
\end{table*}

\begin{figure}[hbt]
\epsfxsize=\hsize
\epsfbox{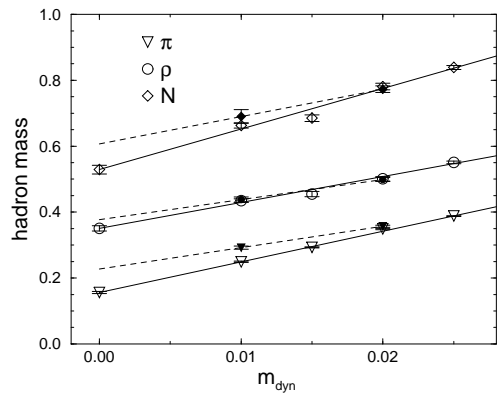}
\caption{The masses for the $\pi$, $\rho$ and $N$ for 2 and 4 flavors.
The unfilled symbols are for 2 flavors and the filled for 4.}
\label{fig:pi_rho_N}
\end{figure}

Figure \ref{fig:pi_rho_N} shows the $\pi$, $\rho$ and nucleon
masses for 2 and 4 flavors as a function of the dynamical quark
mass.  The unfilled symbols are the 2 flavor results; the filled
symbols are for 4 flavors.  The close agreement between the
values for $m_\rho$ is due to our choice of parameters.  The solid
lines are fits to the 2 flavor results and show the extrapolation to
zero quark mass.  The fits have large $\chi^2$ per degree of freedom
(7, 3 and 5 for the $\pi$, $\rho$ and nucleon respectively).  The
reason for this is unknown (finite volume effects are a likely
candidate) and will hopefully be settled by future simulations.
The dashed lines extrapolate the two points we have for 4 flavors.
They are consistent with the larger value for $m_N/m_\rho$ we
found for 4 flavors in the partially quenched hadron spectrum.

\begin{figure}[htb]
\epsfxsize=\hsize
\epsfbox{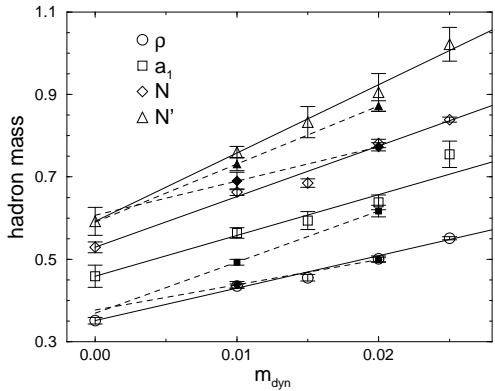}
\caption{Masses for the parity partners, $\rho$ and $a_1$, $N$ and
$N'$.  Unfilled symbols are for 2 flavors, filled circles are for 4
flavors.}
\label{fig:parity_partners}
\end{figure}

Figure \ref{fig:parity_partners} shows the $\rho$, $a_1$, $N$, and
$N'$ masses for 2 and 4 flavors as a function of the dynamical quark
mass.  Again, unfilled symbols are for 2 flavors and filled for 4.
The 2 flavor fits have good $\chi^2$ values for the $a_1$ and $N'$.
It is clear from the figure that the parity partners are not
degenerate in the $m_{\rm dyn}a \rightarrow 0 $ limit.  The dashed
lines extrapolate the 4 flavor results to zero quark mass and one
sees the almost complete degeneracy in this limit.

\begin{figure}[htb]
\epsfxsize=\hsize
\epsfbox{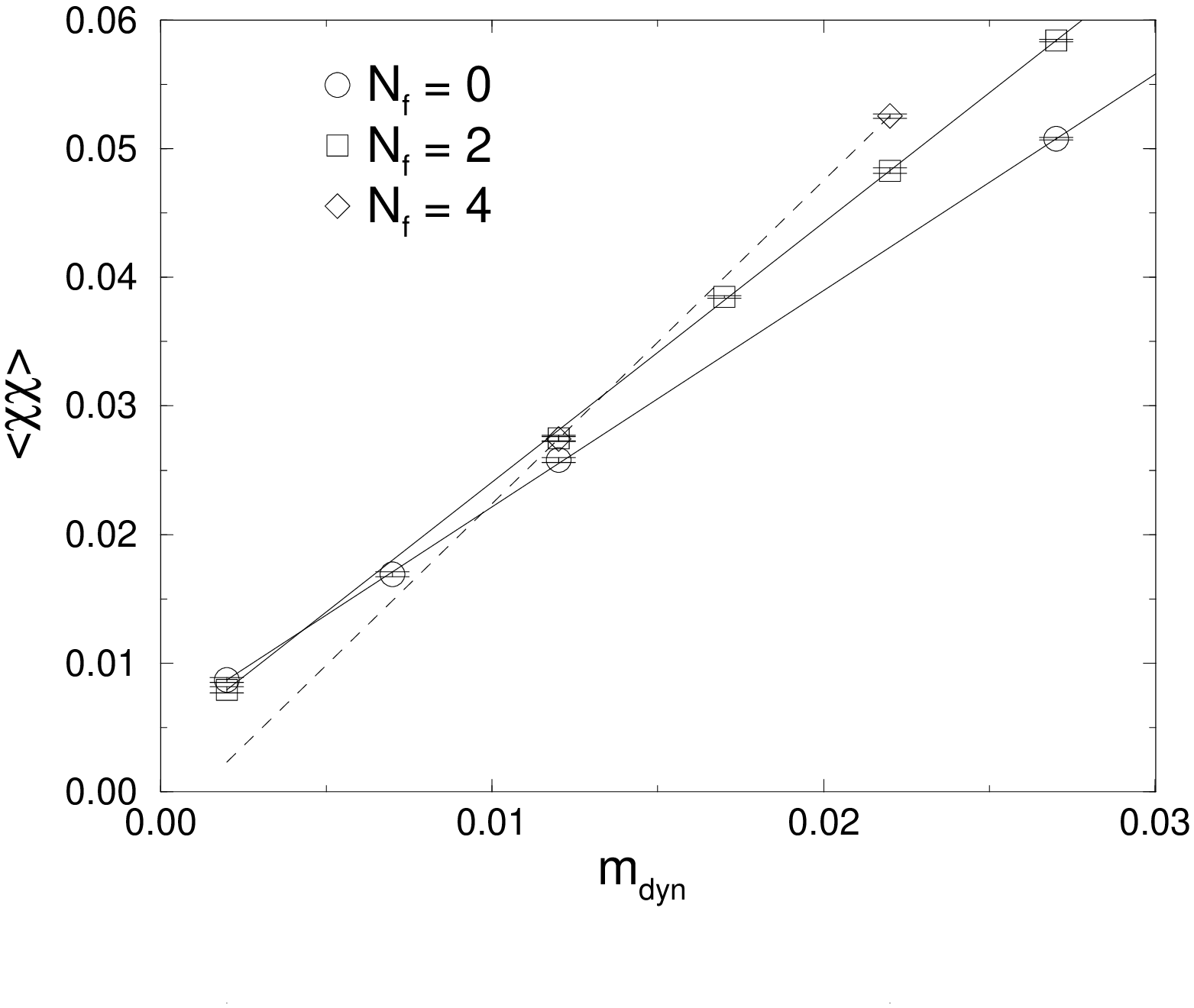}
\caption{The chiral condensate for 0, 2 and 4 flavors.  For the
quenched simulation, the horizontal axis is the valence quark mass.}
\label{fig:chi}
\end{figure}

This apparent degeneracy leads to the question of whether these
lattices are still in a confining, chirally asymmetric phase.  Figure
\ref{fig:chi} shows the staggered fermion chiral condensate, $\langle
\bar{\chi}\chi \rangle $, for 0, 2 and 4 flavors as a function of the
quark mass (the dynamical mass for 2 and 4 flavors).  The solid lines
are fits for 0 and 2 flavors;  the dashed line is an extrapolation
for 4 flavors.  The 4 flavor value at zero quark mass, 0.0023,
is almost 4 times smaller than the 2 flavor value, 0.00854(17).
Using the Gell-Mann--Oakes--Renner formula for $f_\pi$ gives
(using $m_\rho$ at zero quark mass)
\begin{eqnarray}
  {f_\pi}/{m_\rho(0)} = 0.127 & &{\rm 2\;flavors } \\
  {f_\pi}/{m_\rho(0)} = 0.061 & &{\rm 4\;flavors }
\end{eqnarray}
The error on the 2 flavor result is 0.005, but this involves using
fits of poor $\chi^2$ and hence the error is likely unreliable.
However, the difference between the 2 and 4 flavor results is
large on the scale of the error.

\section{CONCLUSIONS}

We have presented data showing a strong $N_f$ dependence in
the hadron spectrum for full QCD for fixed lattice volumes and
spacings (in units of $m_\rho$).  The results are consistent with
the naive expectation that increasing the number of light quarks
suppresses the part of the Dirac eigenvalue spectrum responsible
for chiral symmetry breaking.  Clearly these simulations need to
be extended to larger volumes and weaker couplings, which will
require substantially more computer power.  However, we are
seeing a large effect (almost a factor of 2 change in $f_\pi/m_\rho$)
which is unlikely to be changed drastically by approaching
the continuum limit.

It is of some interest to ask if we have stumbled into a new phase of
QCD, possibly a lattice artifact.  We have measured Wilson lines on our
lattices, in the smaller 16 lattice spacing direction, and find a small
value consistent with a confined phase.  We have no data to support a
more exotic situation, confinement without chiral symmetry breaking.  A
second question is the temperature of our lattices.  For 4 flavors,
estimating from existing data indicates the finite temperature phase
transition should occur on lattices of size $N_t = 10-12$ for $\beta =
5.4$ \cite{nf8}.  Of course our lattices have $N_s = 16 $ and periodic
boundary conditions in the spatial directions, but $N_s = 16$ places
this work in the confined phase.  In addition, the strong first order
nature of the 4 flavor transition means that,  even close to the
critical coupling, the hybrid algorithm will be exploring the other
phase extremely infrequently.

It should be noted that extrapolating our two values for $m_\pi^2$
for 4 flavors does not give $m_\pi^2 \rightarrow 0 $ as the
dynamical quark mass goes to zero.  We have seen this effect in
the partially quenched simulations, where it is consistent with
finite volume effects.  It deserves more study for the full
simulations.  In addition, it would be interesting to see if
partially quenched Wilson spectrum measurements also showed the
effects seen for staggered fermions.

The author would like to thank Jim Sexton, Tony Kennedy and Robert
Edwards for useful discussions and Catalin Malareanu for measuring
the Wilson lines on these lattices.

\end{document}